# How Much Does a Hamiltonian Cycle Weigh?

Hurlee Gonchigdanzan

*Department of Mathematical Sciences*
*University of Wisconsin - Stevens Point*
*Stevens Point, WI 54481*

**1. INTRODUCTION.** A *Hamiltonian cycle* in a graph is a path in the graph which visits each vertex exactly once and returns to the starting vertex. Let $K_n$ be a weighted complete graph with $n$ vertices. We define the weight of an edge as the square of the distance between two end points of the edge. The weight of a path $P$ is the sum of the weights of all edges in the path and denoted by $w(P)$.

The purpose of this article is to investigate how much Hamiltonian cycles weigh in $K_4$ and $K_5$ compare to the total weight of the graph and to establish a precise estimate of it. We have the following results.

**Theorem 1.** *For any Hamiltonian cycle $E$ in $K_4$ we have*

$$\frac{1}{2}w(K_4) \leq w(E) < w(K_4). \tag{1.1}$$

**Theorem 2.** *For any Hamiltonian cycle $E$ in $K_5$ we have*

$$\frac{5-\sqrt{5}}{10}w(K_5) \leq w(E) \leq \frac{5+\sqrt{5}}{10}w(K_5). \tag{1.2}$$

**2. PROOF OF THEOREM 1.** Let $D$ be the complement of the *Hamiltonian cycle $E$*. Suppose that $K_4$, $E$, and $D$ are as shown in Figures 1, 1a and 1b.

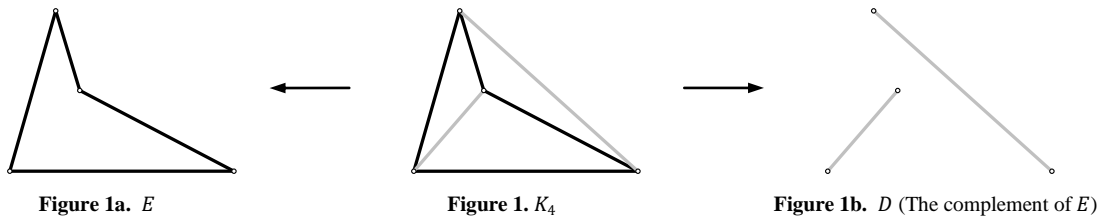

**Figure 1a.** $E$      **Figure 1.** $K_4$      **Figure 1b.** $D$ (The complement of $E$)

Since $w(K_4) = w(D) + w(E)$ and $w(D) > 0$, it immediately follows the right-side inequality of (1.1). On the other hand, the left-side inequality of (1.1) is equivalent to

$$w(D) \leq w(E). \tag{2.1}$$

Therefore we prove (2.1) to complete the proof of Theorem 1. Let $l_k$ and $L_k$ $(1 \leq k \leq 6)$ be the length of a segment that connects a pair of vertices of $K_4$ and the middle point of the segment respectively (See Figures 2a and 2b).

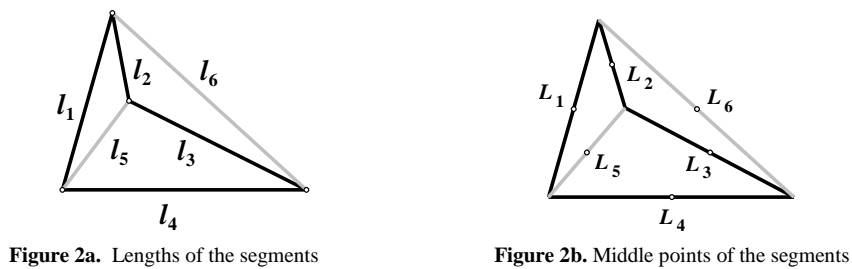

**Figure 2a.** Lengths of the segments      **Figure 2b.** Middle points of the segments

Then $w(D) = l_5^2 + l_6^2$ and $w(E) = l_1^2 + l_2^2 + l_3^2 + l_4^2$. Thus for (2.1), it suffices to prove that

$$l_5^2 + l_6^2 \le l_1^2 + l_2^2 + l_3^2 + l_4^2. \tag{2.2}$$

Observe that regardless of the positions of the four vertices either in space or in the plane, $(L_1L_2L_3L_4)$, $(L_2L_6L_4L_5)$, and $(L_3L_5L_1L_6)$ are parallelograms and

$$\tfrac{1}{2}l_1 = |L_2L_5| = |L_4L_6|, \quad \tfrac{1}{2}l_2 = |L_1L_5| = |L_3L_6|, \quad \tfrac{1}{2}l_3 = |L_2L_6| = |L_4L_5|,$$

$$\tfrac{1}{2}l_4 = |L_1L_6| = |L_3L_5|, \quad \tfrac{1}{2}l_5 = |L_1L_2| = |L_3L_4|, \quad \tfrac{1}{2}l_6 = |L_1L_4| = |L_2L_3|.$$

Recall the parallelogram rule that the sum of the squares of the sides of a parallelogram equals to the sum of the squares of the diagonals. Applying the rule to $(L_1L_2L_3L_4)$, $(L_2L_6L_4L_5)$, and $(L_3L_5L_1L_6)$ we get

$$\tfrac{1}{2}(l_5^2 + l_6^2) = p^2 + q^2, \quad \tfrac{1}{2}(l_1^2 + l_3^2) = q^2 + r^2, \quad \tfrac{1}{2}(l_2^2 + l_4^2) = p^2 + r^2$$

where $p = |L_1L_3|$, $q = |L_2L_4|$, and $r = |L_5L_6|$.

Hence

$$4r^2 + l_5^2 + l_6^2 = l_1^2 + l_2^2 + l_3^2 + l_4^2. \tag{2.3}$$

Since it is possible $r = 0$, (2.3) immediately follows (2.2). ∎

**REMARK.** In fact (2.3) gives us much broader picture of the exact relationship between the sums of the squares of the distances between four points in space or in the plane. It is not just between diagonals and sides.

*Quadrilateral.* (2.3) implies the generalized parallelogram rule (see [**1**], p.22) that the sum of the squares of the sides equals to the sum of the squares of the diagonals and the square of the distance between the middle points of the diagonals (see Figure 3a). Moreover, (2.3) holds when the quadrilateral is concave or complex (see Figures 3b and 3c as in the plane). Some other extensions of the generalized parallelogram can be found in [**2**] and [**3**].

*Tetrahedron.* (2.3) also holds for any tetrahedron (see Figures 3a, 3b, and 3c as in space). In other words, the sum of the squares of any two nonadjacent edges and the square of the distance between the middles points of these two segments equals to the sum of the squares of the other four edges.

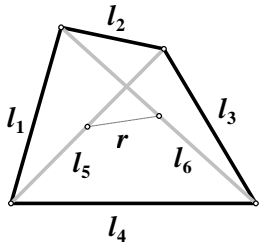
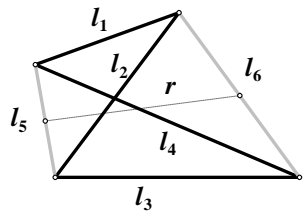
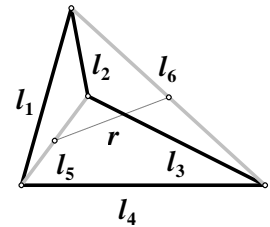

**Figure 3a.** Example 1    **Figure 3b.** Example 2    **Figure 3c.** Example 3

## 3. PROOF OF THEOREM 2.

The idea of the proof is that we construct a sequence of Hamiltonian cycles and obtain the inequality $w(D) < \left(3 - \frac{a_{n-1}}{4a_n}\right) w(E)$ for a certain positive sequence $(a_n)$. Then we take the limit as $n \to \infty$ to get (1.2). The properties of the sequences of $(a_n)$ and the weights $(e_n)$ of the Hamiltonian cycles play essential role in the proof. We prove the following two lemmas first.

**Lemma 1.** *Let $(a_n)_{n \geq 2}$ be a sequence of numbers such that $a_0 = 0, a_1 = 1$, and*

$$a_{n+2} = \frac{12}{16} a_{n+1} - \frac{1}{16} a_n. \tag{3.1}$$

*Then we have*

(i) *The sequence $(a_n)$ is positive and monotone decreasing.*

(ii) $\frac{a_{n+1}}{a_n} > \frac{3+\sqrt{5}}{8}$ *for any $n \geq 1$.*

(iii) $\left(\frac{a_{n+1}}{a_n}\right)$ *is non-increasing.*

(iv) $\lim_{n \to \infty} \frac{a_{n+1}}{a_n} = \frac{3+\sqrt{5}}{8}$.

*Proof of Lemma 1.* For (i), we prove that $a_n \geq 0$ and $a_{n+1} < a_n$ for any $n \geq 1$ by induction. For $n = 1$, we have $a_1 \geq 0$ and $a_2 < a_1$. For the induction step, assuming that $a_n \geq 0$ and $a_{n+1} < a_n$ we need show that $a_{n+1} \geq 0$ and $a_{n+2} < a_{n+1}$. From (3.1) observe that

$$a_{n+2} = \frac{12}{16} a_{n+1} - \frac{1}{16} a_n < \frac{12}{16} a_{n+1} < a_{n+1}$$

and

$$a_{n+2} = \frac{12}{16} a_{n+1} - \frac{1}{16} a_n > \frac{12}{16} a_n - \frac{1}{16} a_n > \frac{1}{16} a_n > 0.$$

Hence we have $a_{n+2} < a_{n+1}$ and $a_{n+2} > 0$ that follows $a_{n+1} > 0$.

For (ii) we use induction. Since $a_2 = 12/16$, (ii) is true for $n = 1$. Now we assume that

$$\frac{a_n}{a_{n-1}} > \frac{3+\sqrt{5}}{8}. \tag{3.2}$$

Using (3.1) and (3.2) we get

$$\frac{a_{n+1}}{a_n} = \frac{12}{16} - \frac{1}{16} \frac{a_{n-1}}{a_n} > \frac{12}{16} - \frac{1}{16} \frac{1}{2(3+\sqrt{5})} = \frac{3+\sqrt{5}}{8}.$$

For (iii), we show that $\frac{a_{n+2}}{a_{n+1}} \geq \frac{a_n}{a_{n+1}}$. Writing (3.1) as

$$\frac{12}{16} - \frac{a_{n+2}}{a_{n+1}} = \frac{1}{16} \frac{a_n}{a_{n+1}},$$

we see that it suffices to show

$$\frac{12}{16} \leq \frac{1}{16} \frac{a_{n+1}}{a_{n+2}} + \frac{a_{n+2}}{a_{n+1}}.$$

Consider the function $f(x) = \frac{1}{16x} + x$ and observe that it is increasing on $[1/4, \infty)$ and $f\left(\frac{3+\sqrt{5}}{8}\right) = \frac{12}{16}$. From (ii) $\frac{a_{n+1}}{a_n} > \frac{3+\sqrt{5}}{8} > \frac{1}{4}$, thus $\frac{1}{16} \frac{a_n}{a_{n+1}} + \frac{a_{n+1}}{a_n} = f\left(\frac{a_{n+1}}{a_n}\right) \geq f\left(\frac{3+\sqrt{5}}{8}\right) = \frac{12}{16}$.

For (iv), first notice that (ii) and (ii) follow that there is a number $c \geq \frac{3+\sqrt{5}}{8}$ such that $\lim_{n\to\infty} \frac{a_{n+1}}{a_n} = c$. Now dividing the both sides of (3.1) by $a_{n+1}$ and then taking the limit as $n \to \infty$ we obtain the equation $\frac{12}{16} - \frac{1}{16c} = c$. Solving the equation, we find $c = \frac{3+\sqrt{5}}{8}$. ∎

**Lemma 2.** *Let $(e_n)_{n\geq 1}$ be a sequence of positive numbers given by $e_{n+2} = \frac{12}{16} e_{n+1} - \frac{1}{16} e_n$. Then $e_n$ can be represented as*
$$e_n = a_n e_2 - b_n e_1$$
*where $(a_n)$ and $(b_n)$ are sequences of positive numbers and $b_{n+1} = \frac{1}{16} a_n$.*

*Proof of Lemma 2.* The proof is by induction on $n$. For $n = 1$, observe that
$$e_3 = \frac{12}{16} e_2 - \frac{1}{16} e_1 \text{ and } e_4 = \left(\frac{12^2}{16^2} - \frac{1}{16}\right) e_2 - \frac{12}{16^2} e_1.$$

For the induction step, we observe that
$$e_{n+2} = \frac{12}{16} e_{n+1} - \frac{1}{16} e_n = \frac{12}{16}(a_{n+1} e_2 - b_{n+1} e_1) - \frac{1}{16}(a_n e_2 - b_n e_1)$$
$$= \left(\frac{12}{16} a_{n+1} - \frac{1}{16} a_n\right) e_2 - \left(\frac{12}{16} b_{n+1} - \frac{1}{16} b_n\right) e_1$$
$$= \left(\frac{12}{16} a_{n+1} - \frac{1}{16} a_n\right) e_2 - \left(\frac{12}{16^2} a_n - \frac{1}{16^2} a_{n-1}\right) e_1.$$

Hence, $(a_n)$ is indeed the sequence given by the following recurrent formula
$$a_{n+2} = \frac{12}{16} a_{n+1} - \frac{1}{16} a_n \tag{3.3}$$
and
$$b_{n+2} = \frac{1}{16}\left(\frac{12}{16} a_n - \frac{1}{16} a_{n-1}\right) = \frac{1}{16} a_{n+1}. \qquad ∎$$

*Proof of Theorem 2:* Let $D$ be the complement of the cycle $E$. Since $w(K_5) = w(D) + w(E)$, (1.2) is equivalent to
$$\frac{3-\sqrt{5}}{2} w(D) \leq w(E) \leq \frac{3+\sqrt{5}}{2} w(D). \tag{3.4}$$

Notice that $D$ is also Hamiltonian. Since $E$ is an arbitrary Hamiltonian cycle in $K_5$, we can easily verify that the right-side inequality of (3.4) follows the left one. Therefore we prove only the right-side inequality of (3.4).

Let $F_1$ be the set of all segments that connect every pair of the vertices of $K_5$ (see Figure 4). Let $|E|_1$ and $|D|_1$ be the sets of the segments that connect the vertices associated with the Hamiltonian cycles $E$ and $D$ (see Figure 4a and 4b).

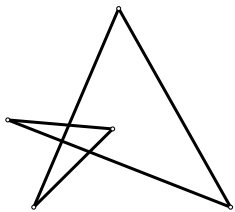 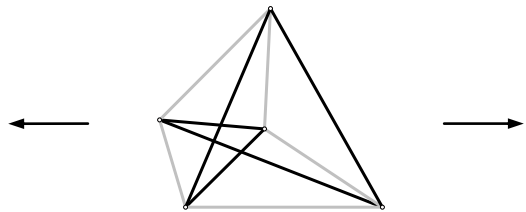 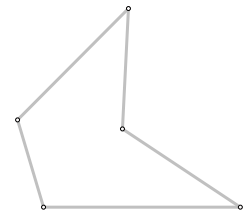

**Figure 4a.** $|E|_1$    **Figure 4.** $F_1$    **Figure 4b.** $|D|_1$

Now, let us consider the middle points of the segments in $|D|_1$ and define $F_2$ as the set of all segments that connect every pair of these five point (see Figure 5). We choose $|D|_2$ as the set of the segments that connect the middle points of the adjacent segments in $|D|_1$ shown in Figure 5b and $|E|_2$ is the complement of $|D|_2$ shown Figure 5a.

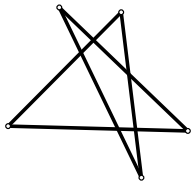
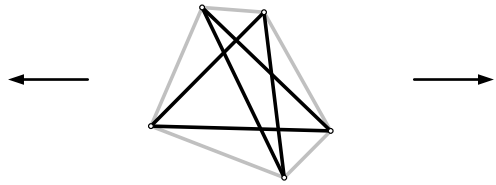
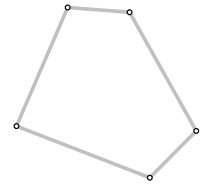

**Figure 5a.** $|E|_2$  **Figure 5.** $F_2$  **Figure 5b.** $|D|_2$

Applying the same idea, we can construct $F_3$, $|E|_3$, and $|D|_3$ in which $|D|_3$ consists of the segments that connect the middle points of the adjacent segments in $|D|_2$ and its complement is defined to be $|E|_3$. If this process is continued we have a sequence of sets of five points and the segments that connect the points:

$$F_1(|E|_1, |D|_1) \to F_2(|E|_2, |D|_2) \to \cdots \to F_n(|E|_n, |D|_n) \to \cdots$$

Denote the sums of the squares of the segments in $|D|_n$ and $|E|_n$ by $d_n$ and $e_n$ respectively. We claim that

$$d_1 + 4e_2 = 3e_1. \tag{3.5}$$

(3.5) is the core of the proof. Let $G(abcd)$ be the set of segments that connect every pair of the vertices $a, b, c$, and $d$. Let us denote the vertices of $E_1$ by $v_1, v_2, v_3, v_4$, and $v_5$ as shown in Figure 6.

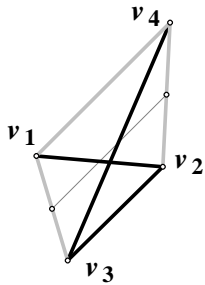
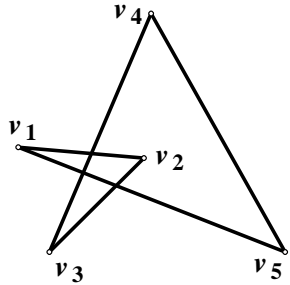
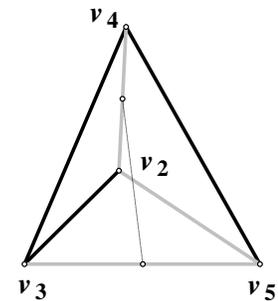

**Figure 6a.** $G(v_1v_2v_3v_4)$  **Figure 6.** $|E|_1$  **Figure 6b.** $G(v_2v_3v_4v_5)$

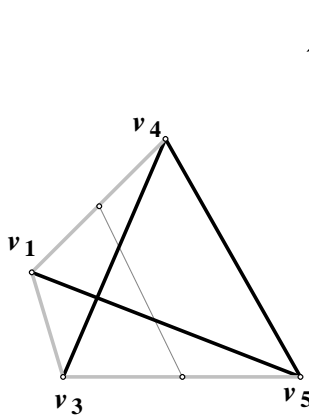
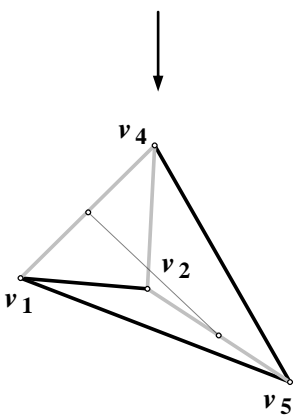
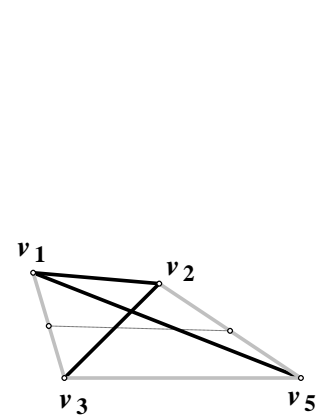

**Figure 6c.** $G(v_3v_4v_5v_1)$  **Figure 6d.** $G(v_4v_5v_1v_2)$  **Figure 6d.** $G(v_5v_1v_2v_3)$

Now using the same idea as (2.3) has been obtained in the proof of Theorem 1, we can have (2.3) for $G(v_1v_2v_3v_4)$, $G(v_2v_3v_4v_5)$, $G(v_3v_4v_5v_1)$, $G(v_4v_5v_1v_2)$, and $G(v_5v_1v_2v_3)$ (see Figure 6a, 6b, 6c, and 6d) where $r$ is the distance between the middle points of the two nonadjacent segments. Combining all of them we can find (3.5). Moreover, by induction $d_n + 4e_{n+1} = 3e_n$. From the way $|E|_n$ and $|D|_{n+1}$ are constructed we have $4d_{n+1} = e_n$ ($n \geq 1$). Hence

$$e_{n+2} = \frac{12}{16}e_{n+1} - \frac{1}{16}e_n.$$

By Lemma 2, (3.5) becomes

$$d_1 + \frac{4}{a_n}e_n = \left(3 - \frac{a_{n-1}}{4a_n}\right)e_1$$

that follows

$$d_1 < \left(3 - \frac{a_{n-1}}{4a_n}\right)e_1. \tag{3.7}$$

By taking limit of (3.7) as $n \to \infty$ and applying Lemma 1(iv) we obtain

$$d_1 < \frac{3 + \sqrt{5}}{2}e_1. \tag{3.8}$$

Equality can indeed hold in (3.8). It can be shown by assigning weights to the edges of $E$ and $D$ in the graph $K_5$ proportional to the squares of the sides and diagonals of a regular pentagon (see Figure 7, 7a, 7b).

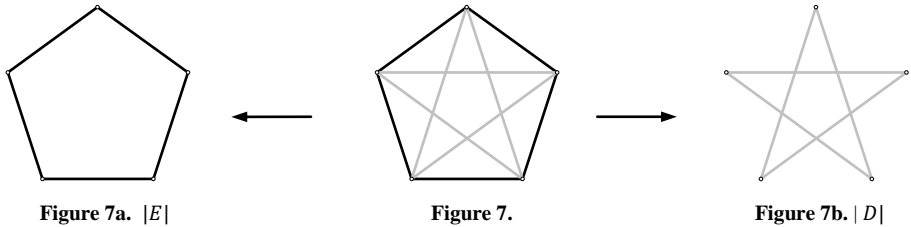

**Figure 7a.** $|E|$       **Figure 7.**       **Figure 7b.** $|D|$

Thus we conclude the right-side inequality of (3.4) ∎